\documentclass[doublecol]{epl2} 
\usepackage{mathtools}

\title{An autonomous and reversible Maxwell's demon}

\author{A. C. Barato \and U. Seifert}

\institute{
  II. Institut f\"ur Theoretische Physik, Universit\"at Stuttgart - Stuttgart 70550, Germany
                    
}

\abstract{ 
Building on a model introduced by Mandal and Jarzynski [Proc. Natl. Acad. Sci. U. S. A., {\bf 109}, (2012) 11641],
we present a simple version of an autonomous reversible
Maxwell's demon. By changing the entropy of a tape consisting
of a sequence of bits passing through the demon,
the demon can lift a mass using the coupling to a heat bath.
Our model becomes reversible by allowing the tape to move in
both directions. In this thermodynamically consistent 
model, total entropy production consists of three terms one
of which recovers the irreversible limit studied by MJ.
Our demon allows an interpretation in terms of an enzyme 
transporting  and transforming molecules between compartments.
Moreover, both genuine equilibrium and a linear 
response regime with corresponding Onsager coefficients are well 
defined. Efficiency and efficiency at maximum power are calculated. 
In linear response, the latter is shown to be bounded by 1/2 if the 
demon operates as a machine and by 1/3 if it is operated as an 
eraser. 
}

\begin{document}

\maketitle

Maxwell's demon is a device which allows to extract work from a heat bath by rectifying fluctuations. It has inspired discussions on the foundations of thermodynamics for more than a century \cite{leff03,maru09}.
Recently, its inherent aspect of information processing in a thermodynamic setting has turned into focus in theoretical studies of specific devices \cite{cao04,andr08,gran11,
vaik11,aver11a,horo11,horo11a,abre11,baue12,espo12b,kish12,mand12},
in the derivation of fluctuation theorems and second law like inequalities 
modified for the presence of information \cite{touc00,touc04,cao09,saga10,horo10,abre11a,saga12,saga12b}, and through pioneering experimental implementations \cite{toya10a,beru12}. Typically, these schemes require the action of a 
macroscopic observer and a {\sl feedback} controller.  

An explicit model for an {\sl autonomous} demon has recently been introduced by Mandal and Jarzynski (MJ) \cite{mand12} for which the relation between mechanical work and information processing, including the 
notoriously subtle issue of the cost of measurement and/or erasure, becomes particular transparent. In the MJ model, 
a frictionless tape composed of bits passes through the demon, 
which is also in contact with a thermal bath and a mechanical reservoir represented by a mass that can go up or down. During a certain time interval the demon interacts with the bit, jumping between its internal states with transition rates
that depend on the mass. Between two interaction time-intervals and ignoring correlations between the bits on the tape, a new incoming bit is instantaneously chosen from the probability distribution corresponding to the ratio of $0$'s and $1$'s
on the tape. The model reaches a periodic steady state, where the outgoing tape is a record of how the internal states of the demon are influenced by the mass. Remarkably, MJ \cite{mand12}
could show that this model fulfills a second law like inequality, with the entropy difference between the incoming and outgoing tape bounding the work that the demon delivers to lift the mass. 
Saturating this equality, however, is arguably not equivalent to equilibrium since the MJ model contains a fundamental asymmetry because the (frictionless) tape is fed into the demon
from one direction, with no option of a spontaneous reversal of directions, as it would happen if the tape was in contact with a thermal reservoir as well.       

The main purpose of this letter is to introduce a model for an autonomous reversible demon whose parameter space contains genuine thermodynamic equilibrium. Inspired by the MJ model, we reformulate and generalize it
as a fully continuous-time stochastic process. Each transition, in principle, is reversible by allowing the tape to move in either direction and not just from left to right as in the MJ model. This system reaches a nonequilibrium
steady state for which the thermodynamic entropy production is easily identified using the framework of stochastic thermodynamics \cite{seif12}. Total entropy production has then three contributions, 
the first corresponding to the difference between the entropy change of the tape and mechanical work as identified by MJ, the second one arising from the fact that the tape can reenter from the right side, and the third one due to
the friction associated with the motion of the tape. 

Genuine equilibrium being part of the phase space of our system, we can develop a linear response treatment with the corresponding identification of the Onsager coefficients. In our analysis, we will
exploit an alternative chemical representation of the demon as an enzyme that transports and transforms particles from one reservoir to another. Another mapping of a demon onto a chemical model has
been suggested independently recently in a somewhat related model \cite{horo12a}.

\begin{figure}
\onefigure[width=70mm]{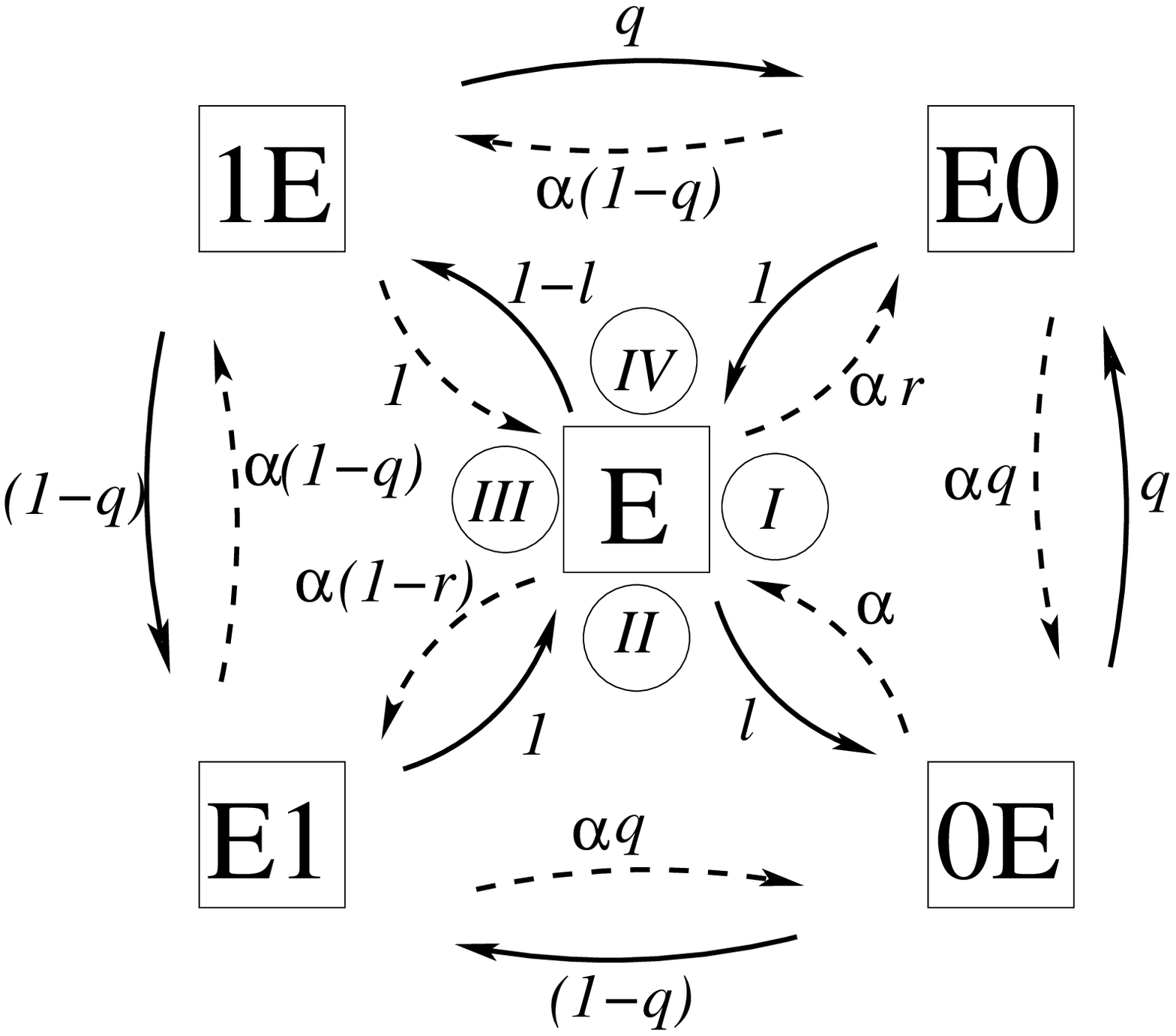}
\caption{The five states model with its transition rates. The full lines represent the transitions where the tape moves from left to right, whereas
the dotted line represent the reverse transitions. All rates in the figure should be multiplied by $k$, which sets the times scale of the transitions. The cycles $I$, $II$, $III$, and $IV$, as defined in equation (\ref{cycles}), are indicated as well.}
\label{fig1}
\end{figure}

The network of five states and the transition rates between them define our model as shown in Fig. \ref{fig1}. First, we consider the case where the tape moves only from left to right, which means that only the transitions represented by the 
full lines in Fig. \ref{fig1}
are possible. In the central state $E$ the demon is not interacting with a bit from the tape. From $E$ the system can either 
jump to $0E$ or $1E$, with rates $kl$ and $k(1-l)$, respectively, where $k$ sets a time scale. These jumps represent a bit coming from the left to interact with
the demon. Therefore, the parameter $l$ is interpreted as the probability of an incoming $0$, while $1-l$ gives the probability of an incoming $1$.  

If the system is at $0E$ or $1E$ it either jumps to $E0$ or $E1$
corresponding to the state of the bit going out to the right.
The transition between the incoming and the outgoing bit arises
from an interaction between the demon and the bit, which may
transform the incoming bit and lift (or lower) the mass,
through an auxiliary process involving two auxiliary states
$E_0$ and $E_1$. An auxiliary transition from $E_0$ to $E_1$ (from
$E_1$ to $E_0$) involves lifting (lowering) the mass by an overall
distance $h$. The transitions between these two auxiliary states
happen on a time scale much faster than $1/k$. Hence, they are
equilibrated on the latter scale leading to a stationary
probability $q$ for being in $E_0$ and $1-q$ for being in $E_1$, where $q=\textrm{e}^f/(1+\textrm{e}^f)$, with $f\equiv mgh/k_BT$ and $T$ is the temperature of the heat bath.
While the initial state $E_b$ ($b=0,1$) for this auxiliary process
is given by the state $bE$ characterizing the incoming bit, the
final state for this auxiliary process $E_{b'}$ ($b'=0,1$) determines the
state of the outgoing bit $Eb'$. The sole purpose
of this auxiliary two-state process is to yield
effective rates for the main network shown in Fig. 1.
We note that this scheme corresponds to the limit where
the interaction time becomes infinite in the MJ model. Moreover, if the transition goes from $0E$ ($1E$) to $E1$ ($E0$) then the mass is lifted (lowered) by a height $h$ and a quantity of work $k_B Tf$ (we set in the following $k_BT=1$)
 will be delivered to (taken from) 
the mechanical reservoir. On the other hand,
if the transition went from $0E$ to $E0$ or from $1E$ to $E1$, then there is no mechanical work involved.

From the states $E0$ and $E1$ the system jumps to the state $E$ with rate $k$. In these transitions, the demon let the transformed bit leave to the right in the outgoing tape and comes back to the central state. Therefore, in a complete cycle starting and 
finishing at the central state $E$, the demon gets a bit from the incoming tape, interacts with this bit, possibly changing it, and then sends the final bit to the outgoing tape.

\begin{figure}
\onefigure[width=70mm]{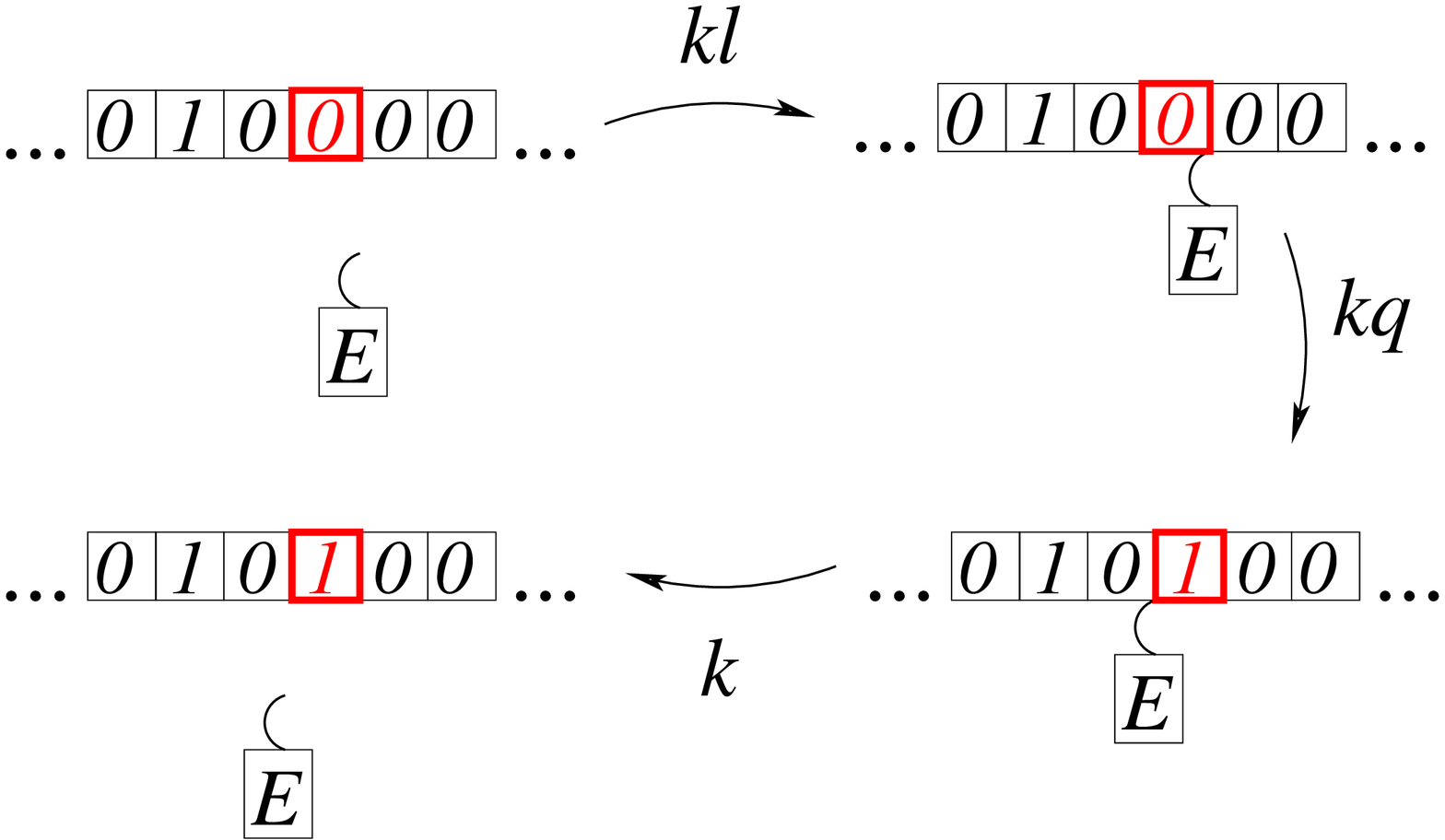}
\caption{The moving demon picture. First the demon binds to a bit $0$ with rate $kl$, it then transforms the $0$ into a $1$ moving to the right with rate $k(1-q)$, and, finally, it detaches from the tape with a rate $k$. Therefore, 
we have a representation of cycle $C_{II}$ defined in (\ref{cycles}).}
\label{fig3}
\end{figure}

We can also consider the model as a demon walking along the tape and changing its bits, see Fig. \ref{fig3}. In this case the demon binds to the tape, interact with the bit, then unbinds from the tape, and proceeds by 
repeating the same process with the next bit.

Summarizing, the model has four cycles, which are
\begin{align} 
&\mathcal{C}_I\equiv (E,0E,E0,E),\qquad\mathcal{C}_{II}\equiv (E,0E,E1,E),\nonumber\\
&\mathcal{C}_{III}\equiv (E,1E,E1,E),\qquad\mathcal{C}_{IV}\equiv (E,1E,E0,E).
\label{cycles}
\end{align}
In the cycle $\mathcal{C}_{II}$ an incoming $0$ is transformed into an outgoing $1$, therefore the demon delivers a quantity of work $f$ to the mechanical reservoir. On the other hand, in the cycle $\mathcal{C}_{IV}$ an incoming $1$ is transformed into
an outgoing $0$ and the demon receives an energy $f$ from the mechanical reservoir. In the other two cycles the incoming bit is not changed and no work is exchanged with the mechanical reservoir. The average 
time for the completion of a cycle is the same for all cycles and given by $3k^{-1}$, which follows from the fact that the escape rate of all states is equal to $k$. Moreover, the four cycles are the only four possible paths that start and finish at $E$, and 
the probability of going through cycle $\mathcal{C}_{II}$ is $l(1-q)$ and through cycle $\mathcal{C}_{IV}$ is $(1-l)q$. Therefore, the rate of work flowing to the mechanical reservoir in the stationary state is
\begin{equation}
\dot{w}_m= k f(l-q)/3.
\end{equation}

The Shannon entropy of the incoming tape is given by
\begin{equation}
H(l)\equiv -l\ln l -(1-l)\ln(1-l).
\end{equation}
This quantity measures how much information the incoming tape contains. Moreover, it is easy to calculate the stationary state probability distribution of the model (see equation (\ref{statprob}) below),
which gives $P_{E0}/P_{E1}=q/(1-q)$, where $P_i$ denotes the 
stationary probability distribution of state $i$. Hence, in the stationary state the entropy of the outgoing tape is $H(q)$ and the rate of information $\dot{h}$ is given by the entropy difference between 
the outgoing tape and the incoming tape divided by the average time to complete a cycle, leading to  
\begin{equation}
\dot{h}= k \Delta H/3,
\end{equation}
where $\Delta H\equiv H(q)-H(l)$.

Remarkably, the rates of mechanical work and information flow follow the inequality  
\begin{equation}
\dot{h}- \dot{w}_m=k (\Delta H-W_m)/3= k D(l||q)/3\ge 0, 
\label{second1}
\end{equation}
where $W_m\equiv (l-q)\ln(q/(1-q)) $ is the average delivered work per cycle and $D(l||q)\equiv l\ln(l/q)+(1- l)\ln((1-l)/(1-q))$ is the Kullback-Leibler distance between the distribution of incoming and outgoing bits,
which is known to be non-negative \cite{cove06}. 
The expressions obtained for $\Delta H$ and $W_m$ are identical to the expressions that are obtained in the MJ model \cite{mand12} in the limit where the interaction time
interval is large. As a first result, we have thus shown that the difference $\Delta H- W_m$ is given by $D(l||q)$. 
We note that in our model the transitions involving a bit coming from the left and the bit leaving to the right are modeled as a Poisson process with a
finite waiting time, whereas in the MJ model these processes are instantaneous.

Depending on the value of $l$ and $q$, the inequality (\ref{second1})
can bound the amount of mechanical work that can be extracted by transferring entropy to the tape or the amount of entropy reduction in the tape generated  by the mechanical energy received from the falling mass \cite{mand12}.
Specifically, without loss of generality we
consider $l\ge 1/2$, corresponding to an excess of $0$ on the incoming tape. For $q>1/2$ and $l>q$ we have $\dot{h}>0$ and $\dot{w}_m>0$,
 hence the demon is a machine that lifts a mass by transferring entropy to the tape. On the other hand, for $q>1/2$ and $l<q$ both quantities change sign and the demon is an 
eraser that uses the energy from the falling mass to erase information (reduce entropy) of the tape. For $q<1/2$ and $l<1-q$ the demon also operates as an eraser, while for $q<1/2$ and $l>1-q$ we have $\dot{h}>0$ and $\dot{w}_m<0$, meaning that the demon
just dissipates energy and does not perform any useful operation.

The above model is a fair description of a frictionless tape interacting with a Maxwell's demon that can lift a mass by exchanging entropy with the tape. 
However, all the reversed transition rates, represented by the dotted lines in Fig. \ref{fig1}, 
are zero and, therefore, entropy production in the sense of stochastic thermodynamics cannot be defined \cite{seif12}. From this perspective, it is then somewhat 
surprising that the fluxes $\dot{h}$ and $\dot{w}_m$ still fulfill the second law like inequality (\ref{second1}).
 In the following, we consider all transition rates in Fig. \ref{fig1} to be non-zero, which means that the tape can also move from right to left.
Such a model has a well defined equilibrium. We would like to compare the standard entropy production with the dissipation rate 
in equation (\ref{second1}). Before considering the full model as a demon interacting with a moving tape we show that the model can also be seen as an enzyme that mediates chemical reactions between different molecules.

The chemical model can be described as follows. Two reservoirs of particles are separated by a membrane. In the left (right) reservoir molecules $0_L$ ($0_R$) and $1_L$ ($1_R$) are dissolved in a solvent. 
These molecules can bind to
 an enzyme $E$ sitting on the membrane that  transforms a molecule of one reservoir into a molecule of the other one. The concentration $c_i$ of the species on the left [right] reservoir are $cl$ [$cr$] for the substance $0_L$ [$0_R$] and $c(1-l)$ 
[$c(1-r)$] for the substance $1_L$ [$1_R$]. Moreover, the concentrations are small ($c\ll 1$) so that the chemical potentials $\mu_{0_L}$, $\mu_{1_L}$, $\mu_{0_R}$ and $\mu_{1_R}$ are related to the concentrations by 
$\mu_i= \ln (2c_i/c)$, where the additive constant $\ln 2$ is chosen for later convenience. 

In the cycle $\mathcal{C}_I$, for example, first a molecule $0_L$ from the left reservoir binds to the enzyme, then the enzyme transforms it into a $0_R$ and, finally, 
 the $0_R$ molecule is released into the right reservoir. 
In the reversed cycle $\overline{\mathcal{C}}_I$, a $0_R$ is taken from the right reservoir, transformed into a $0_L$ by the enzyme that is then released into the left reservoir. The parameter $\alpha$, in Fig. \ref{fig1}, determines the rate
of those transitions that correspond to taking a molecule from the right reservoir and releasing another molecule into the left reservoir: if $\alpha<1$ the left to right transitions are faster than the reversed transitions. Furthermore, 
the enzyme $E$ is connected to a mechanical reservoir, so that when it transforms a $0_L$ ($0_R$) into a $1_R$ ($1_L$), it delivers a quantity $f$ of mechanical work to this reservoir. With the transition rates of Fig. \ref{fig1} the 
escape rates of the four external states are the same, i.e., the average time the enzyme interacts with the molecules is the same for all four molecules. 
     
\begin{table}
\caption{The affinities related to the respective transitions.}
\label{tab1}
\begin{center}
\begin{tabular}{l|c|r}
$m$ & affinity $\mathcal{F}_m$ &  transition\\
\hline
$1$   & $\ln  \alpha^{-1}$ &  all transitions from left to right\\
$2$   & $\mu_{0_L}$  &  $E+0_L\xrightleftharpoons{} 0_LE$\\
$3$   & $\mu_{1_L}$  &  $E+1_L\xrightleftharpoons{} 1_LE$\\
$4$   & $\mu_{0_R}$  &  $E+0_R\xrightleftharpoons{} E0_R$\\
$5$   & $\mu_{1_R}$  &  $E+1_R\xrightleftharpoons{} E1_R$\\
$6$   & $-f$  &  $0_LE \xrightleftharpoons{} E1_R$ and $E0_R \xrightleftharpoons{} 1_LE$ 
\end{tabular}
\end{center}
\end{table}

The model has the following six affinities (or thermodynamic forces): $\ln \alpha^{-1}$, related to the asymmetry between movements from left to right and from right to left, the four chemical potentials $\{\mu_i\}$, and $-f$.
We define the stationary flux between states $i$ and $j$ as $J_{i,j}\equiv P_iw_{ij}-P_jw_{ji}$,
where $w_{ij}$ is the transition rate from $i$ to $j$. Each affinity is related to a flux which enters the entropy production as 
$\dot{s}= \sum_m \mathcal{F}_m J_m$, where $\mathcal{F}_m$ denotes the affinity and $J_m$ the flux between the pair of states related to the affinity. In Table \ref{tab1}, we show the affinities and the transitions they are related to.

\begin{table}
\caption{The distances $d_a^m$ entering equation (\ref{ent1}), where $a$ represents the cycles and $m$ the individual affinities.}
\label{tab2}
\begin{center}
\begin{tabular}{l|cccccr}
cycle  		& $\ln  \alpha^{-1}$	& $\mu_{0_L}$  & $\mu_{1_L}$ & $\mu_{0_R}$ & $\mu_{1_R}$ & $-f$\\
\hline
$C_I$  		& $3$	& $1$  & $0$ & $-1$ & $0$ & $0$\\
$C_{II}$	& $3$	& $1$  & $0$ & $0$ & $-1$ & $1$\\
$C_{III}$  	& $3$	& $0$  & $1$ & $0$ & $-1$ & $0$\\
$C_{IV}$  	& $3$	& $0$  & $1$ & $-1$ & $0$ & $-1$
\end{tabular}
\end{center}
\end{table}

Within stochastic thermodynamics, the entropy production can be written as cycle affinities multiplied by fluxes summed over the fundamental cycles 
as introduced in (\ref{cycles}) \cite{seif12}. In this formulation the entropy production takes the form  
\begin{align}
\dot{s} &=\sum_{a=I}^{IV}J_a\sum_{m=1}^{6}d_a^m\mathcal{F}_m\nonumber\\
&= k(1-\alpha)\ln \alpha^{-1}+\sum_{a=I}^{IV}J_a\sum_{m=2}^{6}d_a^m\mathcal{F}_m.
\label{ent1}
\end{align}
where $d_a^{m}$ is a generalized distance that indicates in how many of the transitions pertaining to the cycle $a$ the affinity $m$ appears (see Table \ref{tab2}). For later convenience, in the second equality 
we have separated the flux $J_1= 1-\alpha$ related to the affinity $\ln \alpha^{-1}$.

Using Table \ref{tab2}, the entropy production becomes 
\begin{align}
\dot{s}=& k(1-\alpha)\ln \alpha^{-1}\nonumber\\
&+J_{0E,E1}(\mu_{0_L}-\mu_{1_R}-f) +J_{1E,E0}(\mu_{1_L}-\mu_{0_R}+f)\nonumber\\ 
&+J_{0E,E0}(\mu_{0_L}-\mu_{0_R}) + J_{1E,E1}(\mu_{1_L}-\mu_{1_R}).
\end{align}

For each external link $a$ we can separate the current through it into two contributions, so that $J_a=J_a^+-J_a^-$. If the link $a$ connects states $i$ and $j$ (from $i$ to $j$), then we have $J_a^+=P_iw_{ij}$ and $J_a^-=P_jw_{ji}$.
Moreover, given a link $a$ that connects states $bE$ and $Eb'$ (with $b,b'$ $0$ or $1$), calculating the stationary state fundamental fluxes we obtain
\begin{align}
J^+_a &= k[X^+_a+\alpha Y^+_a]/[3(1+\alpha)],\nonumber\\
J^-_a &= k\alpha[X^-_a+\alpha Y^-_a]/[3(1+\alpha)],
\label{statprob}
\end{align}
where
\begin{eqnarray}
X^+_a & \equiv & [l+b(1-2l)][q+b'(1-2q)],\nonumber\\
Y^+_a & = & X^-_a\equiv [q+b(1-2q)][q+b'(1-2q)],\nonumber\\
Y^-_a & \equiv &  [q+b(1-2q)][r+b'(1-2r)].
\label{XY}
\end{eqnarray}
With these definitions we can rewrite the entropy production in the instructive form 
\begin{equation}
\frac{\dot{s}}{k}= (1-\alpha)\ln \alpha^{-1}+ \frac{1}{3(1+\alpha)}\sum_{a=I}^{IV}(X_a^+-\alpha^2Y_a^-)\sum_{m=2}^6 d_a^m\mathcal{F}_m,
\end{equation}
which is useful in the following discussion.

Let us now go back to the picture where the demon interacts with the tape. Without loss of generality we consider $\alpha\le 1$, which means that on average the tape moves from left to right. 
As a self-consistency condition, we have to impose  
that the ratio of probabilities of the outgoing bits to the right
is equal the ratio of the incoming bits from the right. This leads to the condition
\begin{equation}
q/(1-q)=r/(1-r),
\label{self}
\end{equation}
i.e., $r= q$. This means that the tape refed from the right has the same statistics as the outgoing tape to the right. With this self-consistency condition and using the definition (\ref{XY}),
we obtain
\begin{equation}
\frac{1}{3(1+\alpha)}\sum_{a}X_a^+\sum_{m=2}^6 d_{am}\mathcal{F}_m=  \frac{1}{3(1+\alpha)}(\Delta H- W_m),
\label{feed}
\end{equation} 
as one contribution to the entropy production. This contribution is related to the difference 
between the statistics of the bits of the tape coming in from the left and the tape going out to the right. It is precisely the term we got in equation (\ref{second1}), obtained there for the case where $\alpha=0$. 
Moreover, from  (\ref{XY}) we also obtain
\begin{equation}
 -\frac{1}{3(1+\alpha)}\alpha^2\sum_{a}Y_a^-\sum_{m=2}^6 d_{am}\mathcal{F}_m= \frac{1}{3(1+\alpha)}\alpha^2D(q||l).
\label{refeed}
\end{equation} 
This term is due to the fact that the tape coming in from the right and tape going out to the left have different probability distributions. In our model, we do not keep correlations between
the bits, considering the simplest approximation where a tape containing a $0$ with probability $q$ is refed from the right to the left, where the probability of a $0$ is $l$. 
Note that for small $\alpha$, where transitions from right to left become rare,
the term (\ref{refeed}) is negligible compared to (\ref{feed}). Combining these terms, the entropy production of the tape model is given by
\begin{equation}
\frac{\dot{s}}{k}=  (1-\alpha)\ln \alpha^{-1}+ \frac{1}{3(1+\alpha)}[\Delta H- W_m+\alpha^2D(q||l)],
\label{tapeentropy}
\end{equation} 
where the first term $(1-\alpha)\ln \alpha^{-1}$ is due to friction between the demon and the tape.

\begin{figure}
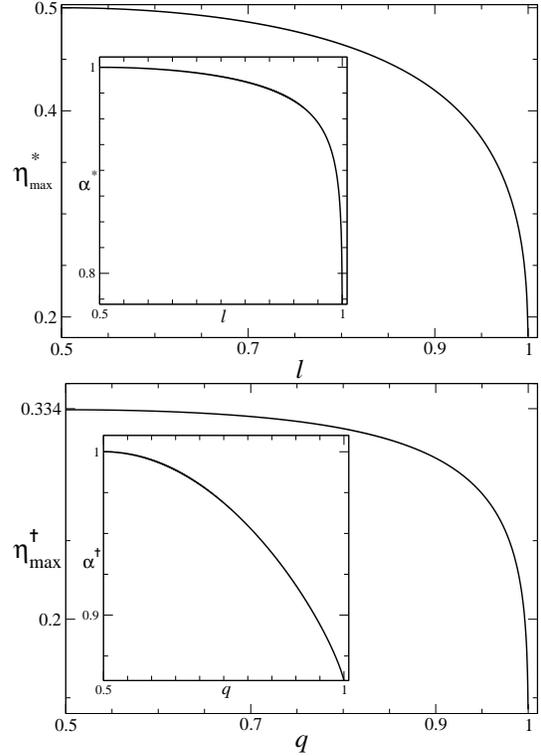

\onefigure[width=70mm]{fig3a.eps}
\onefigure[width=70mm]{fig3b.eps}
\caption{Upper panel: the efficiency  $\eta^*_{\textrm{max}}$ as a function of $l$, when the model operates as a machine. Lower panel: the efficiency $\eta^\dagger_{\textrm{max}}$ as a function of $q$, when 
the model is an eraser. In both cases the optimal $\alpha$ is shown in the inset.}
\label{fig5}
\end{figure}

We now analyze the efficiency of the demon. First we consider the case where $q>1/2$ and $l>q$, so that the demon operates as a machine. If we maximize
the power $\dot{w}_m$ with respect to $q$, i.e. the load $f$, for fixed $l$ and $\alpha$ we find that the maximal power $\dot{w}_m^*$ is reached at $q^*$ given by the solution of the transcendental equation
\begin{equation}
q^* + (1 - q^*) q^* \ln(q^*/(1 - q^*))=l,
\end{equation}
independent of $\alpha$. The efficiency at maximum power is defined as
\begin{equation}
\eta^*(\alpha,l)\equiv \dot{w}_m^*/(\dot{s}^*+\dot{w}_m^*),
\label{eff1}
\end{equation}
where $\dot{s}^*$ is the entropy production at $q^*$. This efficiency depends on $\alpha$ and goes to zero for $\alpha\to 0$ because the dissipation term $(1-\alpha)\ln \alpha^{-1}$ dominates the entropy production. For $\alpha=1$,
the dissipation due to friction is zero, and the dissipation due to the refeeding of the tape becomes more relevant. More precisely, there is an optimal $\alpha^*(l)$, where the efficiency $\eta^*_{\textrm{max}}(l)\equiv\eta^*(\alpha^*(l),l)$ 
is maximal. In Fig. \ref{fig5}, we plot $\alpha^*(l)$, which is always close to $1$ and becomes closer to $1$ as $l$ tends to $1/2$. The efficiency $\eta^*_{\textrm{max}}(l)$
increases as we approach $l=1/2$, where it becomes $1/2$.       

If $q>1/2$ and $l<q$ the demon erases information with a rate $-\dot{h}= -k\Delta H/(3+\alpha)$. If we maximize $-\dot{h}$ with respect to
$l$ for fixed $q$ and $\alpha$, the maximum occurs for $l^{\dagger}=1/2$. The efficiency at maximum "erasure rate" is then 
\begin{equation}
\eta^\dagger(\alpha,q)\equiv -\dot{h}^\dagger/(\dot{s}^\dagger-\dot{h}^\dagger),
\end{equation}
where $\dot{s}^{\dagger}$ is the entropy production for $l^\dagger= 1/2$. Again, as shown in Fig. \ref{fig5}, the optimal $\alpha^\dagger(q)$ is close to $1$ and approaches $1$ for $q\to1/2$. The efficiency $\eta^\dagger_{\textrm{max}}(q)$
at $\alpha^\dagger(q)$ also increases as $q$ approaches $1/2$, where it becomes $1/3$. 

In the final part, we introduce the linear response theory for this demon, for which we  go back to the chemical model. 
In linear response, the stationary fluxes take the form $J_m= \sum_{n=1}^6L_{mn}\mathcal{F}_n$,    
where the affinities $\mathcal{F}$ are given in table \ref{tab1} and the Onsager coefficients $L_{mn}$ are defined as 
\begin{equation}
L_{mn}\equiv \left.\frac{\partial J_m}{\partial \mathcal{F}_n}\right|_{\mathcal{F}=0}.
\end{equation}
Calculating this coefficients we obtain for the matrix $L$  
\begin{equation}
\frac{k}{72}\left(\begin{tabular}{lccccr}
  			
  	$72$	& $12$  & $12$ & $-12$ & $-12$ & $0$\\
		 $12$		& $5$ & $-1$ & $-2$ & $-2$ & $3$\\
		 $12$		& $-1$  & $5$ & $-2$ & $-2$ & $-3$\\
		 $-12$	& $-2$  & $-2$ & $5$ & $-1$ & $3$\\
     	 $-12$	& $-2$  & $-2$ & $-1$ & $5$ & $-3$\\
 $0$		& $3$  & $-3$ & $3$ & $-3$ & $6$
\end{tabular}\right).
\label{onsager}
\end{equation} 
Note that this matrix is symmetric, in agreement with the Onsager reciprocity relations $L_{mn}=L_{nm}$. It is obtained by considering all four chemical potential as independent variables. 
Using the restriction on the chemical potentials, within linear response we have $\mathcal{F}_2=-\mathcal{F}_3$ and $\mathcal{F}_4=-\mathcal{F}_5$. In addition,  
with the self-consistency condition (\ref{self}) we obtain $\mathcal{F}_6=-2 \mathcal{F}_4$. Thus, there are only three independent affinities, which we define as
$\tilde{\mathcal{F}}_1\equiv \mathcal{F}_1$, $\tilde{\mathcal{F}}_2\equiv \mathcal{F}_2$ and $\tilde{\mathcal{F}}_3\equiv \mathcal{F}_6$. Moreover, the associated currents
are $\tilde{J}_1=J_1$, $\tilde{J}_2=J_2-J_3$, and $\tilde{J}_3=J_6$.

The linear response entropy production is given by $\dot{s}= \sum_{m=1}^6\sum_{n=1}^{6} L_{mn}\mathcal{F}_m\mathcal{F}_n$. Using the three independent affinities
it can be rewritten as $\dot{s}= \sum_{m=1}^3\sum_{n=1}^{3} \tilde{L}_{mn}\tilde{\mathcal{F}}_m\tilde{\mathcal{F}}_n$, where the the matrix 
$\tilde{L}_{mn}\equiv \left.\frac{\partial \tilde{J}_m}{\partial \tilde{\mathcal{F}}_n}\right|_{\tilde{\mathcal{F}}=0}$ is given by
\begin{equation}
\frac{k}{72}\left(\begin{tabular}{lcr}
  			
 	$72$	&  $0$ & $0$\\
	 $0$	&  $12$ & $6$\\
	 $0$	&  $6$  & $3$ 
		 
\end{tabular}\right).
\label{onsager2}
\end{equation} 
Note that in order to calculate this matrix the relations between the chemical potentials and the self-consistency condition have to be implemented before taking the derivatives.
The explicit form of the entropy production is then
\begin{align}
\dot{s}/k & =(\ln \alpha^{-1})^2+ (2/3)(l-1/2)[(l-1/2)-(q-1/2)]\nonumber\\
& -(2/3)(q-1/2)[(l-1/2)-(q-1/2)].
\label{entropylinear}
\end{align}
The second term in the right hand side is given by $\tilde{\mathcal{F}}_2\tilde{J}_2$, and is related to the dissipation term $\frac{k}{3(1+\alpha)}[\Delta H+\alpha^2D(q||l)]$ in (\ref{tapeentropy}), 
whereas the third term is $\tilde{\mathcal{F}}_3\tilde{J}_3$, which corresponds to the
mechanical work. 
 
Using these results we can prove bounds on the efficiency at maximum power in the linear response regime.
For $q>1/2$ and $l>q$, where the demon operates as a machine, the power for fixed $l$ and $\alpha$ is maximal at $q^*=(l+1/2)/2$. 
Calculating the power and the entropy production at $q^*$, we obtain with the definition (\ref{eff1}) the following efficiency at maximum power
\begin{equation}
\eta^*(\alpha,l)= \frac{1}{2}\left(\frac{1}{3[\ln(\alpha^{-1})/(l-1/2)]^2+1}\right)\le\frac{1}{2}.
\end{equation}
When the demon operates as an eraser, noting that within linear response $\Delta H= 2[(l-1/2)^2-(q-1/2)^2]$, we obtain that
the erasure rate for fixed $q$ and $\alpha$ is maximal at $l^\dagger=1/2$. Calculating the linear response theory entropy production at $l^\dagger$ we obtain
\begin{equation}
\eta^\dagger(\alpha,q)= \frac{1}{3}\left(\frac{1}{[\ln(\alpha^{-1})/(q-1/2)]^2+1}\right)\le\frac{1}{3}.
\end{equation}
Hence, when the demon operates as a machine the efficiency at maximum power is bounded by the well known Curzon-Ahlborn result $1/2$, valid for strongly coupled machines \cite{seif12,espo09}. However,
when the demon operates as an eraser the efficiency at maximum erasure rate is bounded by $1/3$. 

In conclusion, our stochastic fully time-continuous version 
of a Maxwell's demon contains 
thermal and mechanical equilibrium as a possible state. It thus allows
to develop a systematic linear response theory and to identify Onsager
coefficients. As for ordinary machines, we find that efficiency at maximum
power is bounded by $1/2$. Operating as an eraser, however, the efficiency at 
maximum erasure rate  is bounded by $1/3$. These linear response
results can be extended to the non-linear regime where we recover in the
irreversible limit the inequality between mechanical power and information
flow found in \cite{mand12}. The derivation of our results was facilitated by 
exploiting an analogy between this demon and an enzyme transforming and
transporting molecules between two compartments. Exploring whether and how these
results can contribute to  a unified theory of autonomous machines that includes the aspect of 
information processing at finite temperature explicitly remains a 
challenging perspective for future work.

\acknowledgments
Support by the ESF though the network EPSD  is gratefully acknowledged. We thank D. Abreu for carefully reading the paper.


\begin{thebibliography}{10}
\expandafter\ifx\csname url\endcsname\relax\def\url#1{\texttt{#1}}\fi

\bibitem{leff03}
\Name{Leff H.~S. \and Rex A.~F.} \Book{{M}axwell's Demon : Entropy, Classical
  and Quantum Information, Computing} (IOP, Bristol and Philadelphia) 2003.

\bibitem{maru09}
\Name{Maruyama K., Nori F. \and Vedral V.} \REVIEW{Rev. Mod.
  Phys.}{81}{2009}{1}.

\bibitem{cao04}
\Name{Cao F.~J., Dinis L. \and Parrondo J. M.~R.} \REVIEW{Phys.\ Rev.\
  Lett.}{93}{2004}{040603}.

\bibitem{andr08}
\Name{Andrieux D. \and Gaspard P.} \REVIEW{Proc.\ Natl.\ Acad.\ Sci.\
  U.S.A.}{105}{2008}{9516}.

\bibitem{gran11}
\Name{Granger L. \and Kantz H.} \REVIEW{Phys. Rev. E}{84}{2011}{061110}.

\bibitem{vaik11}
\Name{Vaikuntanathan S. \and {J}arzynski C.} \REVIEW{J.\ Chem.\
  Phys.}{134}{2011}{054107}.

\bibitem{aver11a}
\Name{Averin D.~V., M\"ott\"onen M. \and Pekola J.~P.} \REVIEW{Phys.\ Rev.\
  B}{84}{2011}{245448}.

\bibitem{horo11}
\Name{Horowitz J.~M. \and Parrondo J. M.~R.} \REVIEW{EPL}{95}{2011}{10005}.

\bibitem{horo11a}
\Name{Horowitz J.~M. \and Parrondo J. M.~R.} \REVIEW{New J.
  Phys.}{13}{2011}{123019}.

\bibitem{abre11}
\Name{Abreu D. \and Seifert U.} \REVIEW{EPL}{94}{2011}{10001}.

\bibitem{baue12}
\Name{Bauer M., Abreu D. \and Seifert U.} \REVIEW{J. Phys. A Math.
  Theor.}{45}{2012}{162001}.

\bibitem{espo12b}
\Name{Esposito M. \and Schaller G.} \REVIEW{EPL}{99}{2012}{30003}.

\bibitem{kish12}
\Name{Kish L.~B. \and Granqvist C.~G.} \REVIEW{EPL}{98}{2012}{68001}.

\bibitem{mand12}
\Name{Mandal D. \and Jarzynski C.} \REVIEW{Proc.\ Natl.\ Acad.\ Sci.\
  U.S.A.}{109}{2012}{11641}.

\bibitem{touc00}
\Name{Touchette H. \and Lloyd S.} \REVIEW{Phys.\ Rev.\ Lett.}{84}{2000}{1156}.

\bibitem{touc04}
\Name{Touchette H. \and Lloyd S.} \REVIEW{Physica A}{331}{2004}{140}.

\bibitem{cao09}
\Name{Cao F.~J. \and Feito M.} \REVIEW{Phys. Rev. E}{79}{2009}{041118}.

\bibitem{saga10}
\Name{Sagawa T. \and Ueda M.} \REVIEW{Phys. Rev. Lett.}{104}{2010}{090602}.

\bibitem{horo10}
\Name{Horowitz J.~M. \and Vaikuntanathan S.} \REVIEW{Phys. Rev.
  E}{82}{2010}{061120}.

\bibitem{abre11a}
\Name{Abreu D. \and Seifert U.} \REVIEW{Phys.\ Rev.\ Lett.}{108}{2012}{030601}.

\bibitem{saga12}
\Name{Sagawa T. \and Ueda M.} \REVIEW{Phys. Rev. E}{85}{2012}{021104}.

\bibitem{saga12b}
\Name{Sagawa T. \and Ueda M.} \REVIEW{Phys.\ Rev.\ Lett.}{109}{2012}{180602}.

\bibitem{toya10a}
\Name{Toyabe S., Sagawa T., Ueda M., Muneyuki E. \and Sano M.} \REVIEW{Nature
  Phys.}{6}{2010}{988}.

\bibitem{beru12}
\Name{B{\'e}rut A., Arakelyan A., Petrosyan A., Ciliberto S., Dillenschneider
  R. \and Lutz E.} \REVIEW{Nature}{483}{2012}{187}.

\bibitem{seif12}
\Name{Seifert U.} \REVIEW{Rep. Prog. Phys.}{75}{2012}{126001}.

\bibitem{horo12a}
\Name{Horowitz J., Sagawa T. \and Parrondo J. M.~R.}
  \REVIEW{arXiv:1210.6448}{}{2012}.

\bibitem{cove06}
\Name{Cover T.~M. \and Thomas J.~A.} \Book{Elements of information theory}
   (Wiley, Hoboken, NJ, and Canada)
  2006.

\bibitem{espo09}
\Name{Esposito M., Lindenberg K. \and van~den Broeck C.} \REVIEW{Phys. Rev.
  Lett.}{102}{2009}{130602}.

\end{thebibliography}

\end{document}